\documentclass[iop]{emulateapj}

\def\ltsima{$\; \buildrel < \over \sim \;$}
\def\simlt{\lower.5ex\hbox{\ltsima}}
\def\gtsima{$\; \buildrel > \over \sim \;$}
\def\simgt{\lower.5ex\hbox{\gtsima}}
\def\kms{{\rm\,km\,s^{-1}}}

\def\kpc{{\rm\,kpc}}

\def\pc{{\rm\,pc}}

\def\deg{^\circ}

\def\s{\ifmmode \widetilde \else \~\fi}
\def\={\overline}

\def\spose#1{\hbox to 0pt{#1\hss}}

\def\lta{\mathrel{\spose{\lower 3pt\hbox{$\mathchar"218$}}
     \raise 2.0pt\hbox{$\mathchar"13C$}}}
\def\gta{\mathrel{\spose{\lower 3pt\hbox{$\mathchar"218$}}
     \raise 2.0pt\hbox{$\mathchar"13E$}}}
\def\Dt{\spose{\raise 1.5ex\hbox{\hskip3pt$\mathchar"201$}}}    
\def\dt{\spose{\raise 1.0ex\hbox{\hskip2pt$\mathchar"201$}}}    

\def\dotsfill{\leaders\hbox to 1em{\hss.\hss}\hfill}

\def\FeH{{\rm[Fe/H]}}

\shorttitle{The Perseus~I dwarf galaxy}
\shortauthors{N. F. Martin et al.}

\begin{document}


\title{Perseus~I: A distant satellite dwarf galaxy of Andromeda}


\author{Nicolas F. Martin$^{1,2}$, Edward F. Schlafly$^2$, Colin T. Slater$^{3}$, Edouard J. Bernard$^4$, Hans-Walter Rix$^2$, Eric F. Bell$^3$, Annette M. N. Ferguson$^4$, Douglas P. Finkbeiner$^5$, Benjamin P. M. Laevens$^{1,2}$, William S. Burgett$^6$, Kenneth C. Chambers$^6$, Peter W. Draper$^7$, Klaus W. Hodapp$^6$, Nicholas Kaiser$^6$, Rolf-Peter Kudritzki$^6$, Eugene A. Magnier$^6$, Nigel Metcalfe$^7$, Jeffrey S. Morgan$^6$, Paul A. Price$^8$, John L. Tonry$^6$, Richard J. Wainscoat$^6$, Christopher Waters$^6$}
\email{nicolas.martin@astro.unistra.fr}

\altaffiltext{1}{Observatoire astronomique de Strasbourg, Universit\'e de Strasbourg, CNRS, UMR 7550, 11 rue de l'Universit\'e, F-67000 Strasbourg, France}
\altaffiltext{2}{Max-Planck-Institut f\"ur Astronomie, K\"onigstuhl 17, D-69117 Heidelberg, Germany}
\altaffiltext{3}{Department of Astronomy, University of Michigan, 500 Church St., Ann Arbor, MI 48109}
\altaffiltext{4}{Institute for Astronomy, University of Edinburgh, Royal Observatory, Blackford Hill, Edinburgh EH9 3HJ, UK}
\altaffiltext{5}{Harvard-Smithsonian Center for Astrophysics, 60 Garden Street, Cambridge, MA 02138, USA}
\altaffiltext{6}{Institute for Astronomy, University of Hawaii at Manoa, Honolulu, HI 96822, USA}
\altaffiltext{7}{Department of Physics, Durham University, South Road, Durham DH1 3LE, UK}
\altaffiltext{8}{Department of Astrophysical Sciences, Princeton University, Princeton, NJ 08544, USA}

\begin{abstract}
We present the discovery of a new dwarf galaxy, Perseus~I/Andromeda~XXXIII, found in the vicinity of Andromeda (M31) in stacked imaging data from the Pan-STARRS1 $3\pi$ survey. Located $27.9\deg$ away from M31, Perseus~I has a heliocentric distance of $785\pm65\kpc$, compatible with it being a satellite of M31 at $374^{+14}_{-10}\kpc$ from its host. The properties of Perseus~I are typical for a reasonably bright dwarf galaxy ($M_V = -10.3\pm0.7$), with an exponential half-light radius of $r_h = 1.7\pm0.4$ arcminutes or $r_h = 400^{+105}_{-85}\pc$ at this distance, and a moderate ellipticity ($\epsilon = 0.43^{+0.15}_{-0.17}$). The late discovery of Perseus~I is due to its fairly low surface brightness ($\mu_0=25.7^{+1.0}_{-0.9}$~mag/arcsec$^2$), and to the previous lack of deep, high quality photometric data in this region. If confirmed to be a companion of M31, the location of Perseus~I, far east from its host, could place interesting constraints on the bulk motion of the satellite system of M31.
\end{abstract}

\keywords{Local Group --- galaxies: individual: Per~I --- galaxies: individual: And~XXXIII}

\section{Introduction}

Aside from the individual importance of dwarf galaxies to constrain low-mass galaxy formation in a cosmological context, these faint systems are powerful probes of their environment. Not only can they be used to study the intricacies of star formation processes in their comparatively shallow potentials \citep[e.g.][]{brown12} or constrain the dark matter content of their host \citep[e.g.][]{watkins10}, but they can also provide information on the transverse bulk motion of external satellite systems with an accuracy that is hard, or even impossible to reach through direct proper motion measurements \citep{vandermarel08}. In this context, satellite dwarf galaxies at large projected distances from their hosts are particularly valuable. However, they also prove the most difficult to find as they require the surveying of larger and larger swaths of the sky at increasing distances from their hosts.

Our view of the satellite system of the Andromeda galaxy (M31) has been transformed over the last decade through systematic observations of its surrounding regions. The Pan-Andromeda Archeological Survey (PAndAS; \citealt{mcconnachie09}) unveiled many new companion dwarf galaxies of M31 \citep[e.g.][]{martin13b}, yet the survey limits confine these discoveries within $\sim150\kpc$, or $\sim11\deg$, from M31 in projection. In parallel, the latest releases of the Sloan Digital Sky Survey (SDSS, DR8 and beyond; \citealt{aihara11}) include the coverage of large regions south of M31, which led to the discovery of its two distant satellite dwarf galaxies, And~XXVIII and And~XXIX, located at $27.7\deg$ and $15.1\deg$, respectively \citep{bell11,slater11}. But it is now the $3\pi$ survey conducted with the first telescope of the Panoramic Survey and Rapid Response Systen (Pan-STARRS1, or PS1 for short; \citealt{kaiser10}) that currently holds the best promise for the discovery of distant companions of M31. Conducted in optical and near-infrared bands, this survey is effectively not limited in its coverage around M31, thereby opening the possibility to efficiently uncover reasonably bright ($M_V\simlt-8.5$) dwarf galaxies beyond the virial radius of their host. Already, PS1 has enabled us to find two bright dwarf galaxies, Lacerta~I and Cassiopeia~III \citep{martin13a}, with the former located at $20.3\deg$ from M31.

In this letter, we present the discovery of a new system found in the PS1 $3\pi$ imaging data. Located at $27.9\deg$ from M31 on the sky, it appears to be the satellite most distant from M31 in projection. The new dwarf galaxy is located in the Perseus constellation, but is also likely a satellite of M31, so we follow the dual-naming convention emphasized in \citet{martin09} to avoid ambiguity and refer to it as Perseus~I (Per~I), or Andromeda~XXXIII. The structure of the letter is as follows: Section~2 focuses on the PS1 survey and the data used for this work, Section~3 presents the derivation of the new dwarf galaxy's properties, and we finally discuss the implications of this discovery in Section~4

\section{The Pan-STARRS1 survey}
The PS1 $3\pi$ survey (Chambers et al., in preparation) is conducting an imaging survey of the sky north of declination $-30\deg$. The dedicated PS1 1.8-meter telescope observes the sky in five photometric bands ranging from the blue to the near infrared ($g_\mathrm{P1}r_\mathrm{P1}i_\mathrm{P1}z_\mathrm{P1}y_\mathrm{P1}$; \citealt{tonry12}) with a wide-field, 1.4-Gpixel camera which has a total field of view of 3.3 sq. degrees. This ensures a rapid scanning of the sky with four planned observations per year and per filter. Each of these images has an exposure time of $43/40/45/30/30$ seconds in the $g_\mathrm{P1}/r_\mathrm{P1}/i_\mathrm{P1}/z_\mathrm{P1}/y_\mathrm{P1}$ bands, respectively \citep{metcalfe13}, with a median image quality of $1.27/1.16/1.11/1.06/1.01$ arcseconds for these bands. Once the individual frames are downloaded from the summit, they are automatically processed with the Image Processing Pipeline \citep{magnier06,magnier07,magnier08} to generate a photometric catalogue.

At the time of this letter, the region around M31 had been observed for three consecutive seasons starting in May 2010, resulting in a total of up to twelve exposures per band. However, weather, chip gaps, and technical problems translate into a varying but lower number of exposures available for stacking at any location in the survey. In the region around the new dwarf galaxy, this translates into 10-$\sigma$ completeness limits of 22.2/22.0 in the $r_\mathrm{P1}$/$i_\mathrm{P1}$ bands, respectively, and a coverage factor close to unity.

In order to optimize the photometric catalogue around the new dwarf galaxy (see E. Bernard et al., in preparation, for more details), we combine all the individual images from all bands in this region to produce the deepest possible image from which sources are detected. Forced photometry is then performed at the location of these detections on all individual images. The catalogues thereby produced are averaged for a given detection and band, before being calibrated onto the PS1 photometric system via the single exposure PS1 catalogue \citep{schlafly12}. The photometry is performed with the {\sc daophot/allstar/allframe} suite of programs \citep{stetson94}, and only objects with uncertainties lower than 0.2 in both the $r_\mathrm{P1}$ and the $i_\mathrm{P1}$ band are kept. An additional cut on the absolute value of the sharpness parameter ($<2$) and a magnitude-dependent cut on $\chi$ ensures that the detections used below have the properties expected from point sources.

In what follows, the distance to M31 is assumed to be $779^{+19}_{-18}\kpc$ \citep{aconn12} and, if necessary, we model the uncertainties on this value by drawing directly from the posterior probability distribution function provided by these authors. All magnitudes are dereddened using the \citet{schlegel98} maps, and the following extinction coefficients : $A_{g_{P1}}/E(B-V) = 3.172$, $A_{r_{P1}}/E(B-V) = 2.271$, and $A_{i_{P1}}/E(B-V) = 1.682$ \citep{schlafly11}.

\section{Perseus~I}
In the six months following the discovery of Lac~I and Cas~III as overdensities of candidate red giant branch (RGB) stars at the distance of M31, the catalogue of stellar detections in the stacked images of the $3\pi$ survey of PS1 has been cleaned of multiple repeats. These stemmed from the independent  ingestion in the database of the overlapping regions of contiguous $0.5\deg\times0.5\deg$ sky cells used to locally stack images. Now that these spurious objects are removed from the global catalogue of stacked sources, the source density on the sky is much smoother, enabling search algorithms to uncover fainter systems than the relatively bright Lac~I and Cas~III.

As a first step, we produce a fine-scale density map of M31 RGB-like stars within a region of $\pm40\deg$ east/west and north/south from M31, with pixels of $1.5'$ on the side, which we then smooth with a Gaussian kernel of dispersion $3'$. At the distance of M31, this kernel corresponds to $\sim680\pc$, typical of the size of its dwarf galaxies \citep{brasseur11b}. The resulting map shows only a few significant detections above the background level, as expected, many of which correspond to known M31 dwarf galaxies. These can be as faint as And~XXIV ($M_V=-7.6\pm0.5$; \citealt{richardson11}) when the PS1 observing conditions are favorable. Some of these detections correspond to either foreground or background astrophysical objects, or to a few artifacts in the data, but a handful of them have the properties expected for candidate M31 dwarf galaxies. While most of these require follow-up observations to confirm their nature or reliably determine their properties, the detection centered on ICRS $(\alpha,\delta)=(3^\mathrm{h}01^\mathrm{m}23.6^\mathrm{s},+40\deg59'18'')$, which has no counterpart in NED\footnote{http://ned.ipac.caltech.edu} or Simbad\footnote{http://simbad.u-strasbg.fr/simbad/}, is significant enough to yield the unambiguous detection of the new dwarf galaxy Perseus~I/Andromeda~XXXIII.

\begin{figure*}
\begin{center}
\includegraphics[width=0.45\hsize,angle=270]{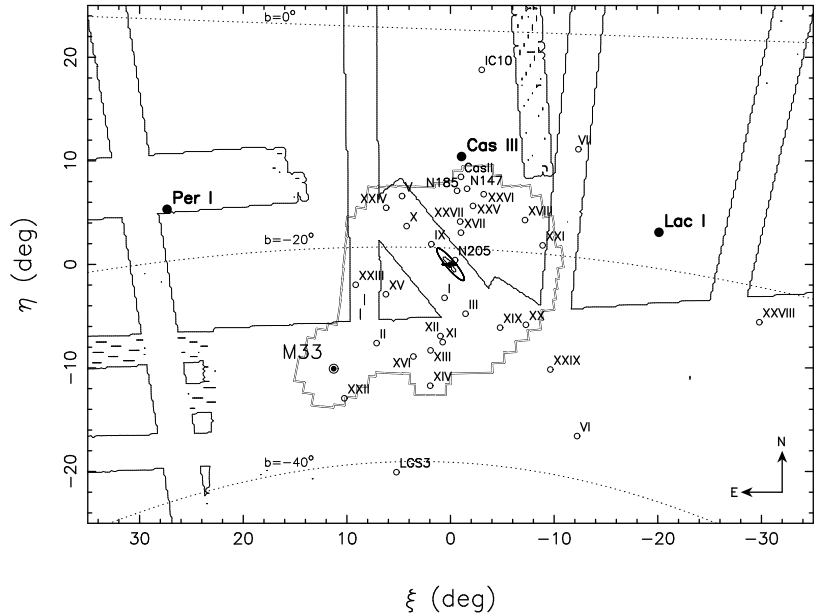}
\caption{\label{map} Distribution of dwarf galaxies around M31 and M33, projected on the tangential plane centered on M31. Known satellites are shown as open circles while filled circles locate the three dwarf galaxies uncovered within the PS1 $3\pi$ data. The two polygons delimit the SDSS and PAndAS surveys (thin line and double line, respectively), and the dotted lines represent lines of Galactic latitude $0\deg$, $-20\deg$, and $-40\deg$. Per~I does overlap with the SDSS footprint, but the SDSS data in this region are of very poor quality.}
\end{center}
\end{figure*}

As shown in Figure~\ref{map}, the location of Per~I with respect to M31 is quite extreme, $27.9\deg$ to the east. But we show below that its distance places it within the sphere of influence of M31. Whether it is truly bound to M31 or not needs to be confirmed from its systemic radial velocity but, if confirmed, Per~I would be, along with And~XXVIII, the furthest dwarf galaxy from M31 in projection. The figure also highlights that Per~I is in fact within the SDSS footprint. However, it resides at the very edge of the survey and, in addition, this region was observed under very poor conditions. The combination of these two drawbacks led to Per~I escaping searches of M31 satellites in SDSS DR8 \citep[e.g.][]{slater11}. However, the SDSS images do show a feeble overdensity of sources at the location of Per~I.

\begin{figure}
\begin{center}
\includegraphics[width=0.7\hsize,angle=270]{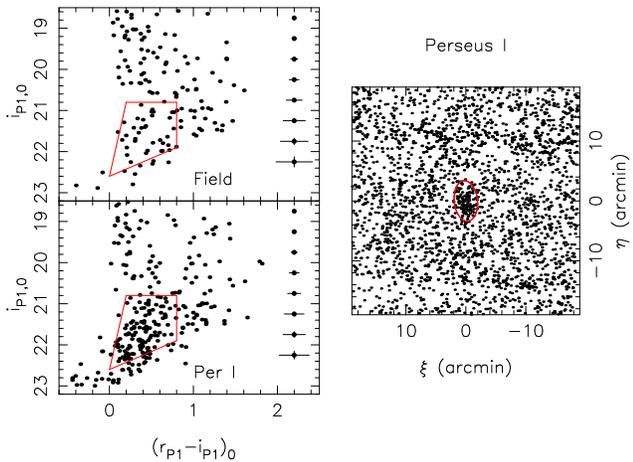}
\hspace{1cm}
\caption{\label{CMD_map}CMD of stars within two half-light radii of the center of Per~I (bottom-left) and that of star within a similarly sized region, $20'$ to the east of Per~I (top-left). Average photometric uncertainities are represented by the error bars to the right of the CMDs. The overdensity of stars in the dwarf galaxy's CMD is typical of bright RGB stars at the distance of M31. Stars specifically selected within the red polygon produce the spatial overdensity apparent on the map of the right panel. The red ellipse indicates the region within two half-light radii of the dwarf galaxy's centroid, as measured in Section~\ref{structure}. Data artifacts stemming from depth differences produce the linear overdensities visible at $\eta\sim10\deg$ and $\eta\sim-12\deg$.}
\end{center}
\end{figure}

\begin{figure}
\begin{center}
\includegraphics[width=0.7\hsize,angle=270]{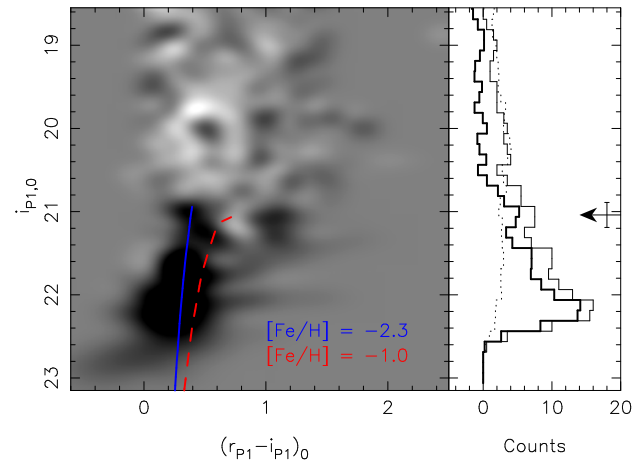}
\caption{\label{LF}The left panel displays a field-subtracted, smoothed CMD of sources within three half-light radii of Per~I (the smoothing kernel has a dispersion of 0.1 magnitude along both the color and magnitude axes). The overdensity of RGB Per~I stars is clearly visible and consistent with \textsc{Parsec} isochrones \citep{bressan12} of age 12.7~Gyr and metallicity $\FeH = -2.3$ (blue, full line), or $\FeH=-1.0$ (red, dashed line), moved to the favored distance estimate for Per~I ($785\kpc$, $m-M = 24.49$). The luminosity function of stars within two half-light radii of Per~I is shown in the panel on the right as a thin line, and compared to the luminosity function of a larger field region scaled to the same area (dotted line). The field-corrected luminosity function of Per~I is represented by the thick line. The arrow marks the TRGB of the dwarf galaxy derived in Section~\ref{distance}, and the vertical error bar attached to it represents the uncertainty on this value.}
\end{center}
\end{figure}

Figure~\ref{CMD_map} displays the color-magnitude diagram (CMD) of stars within two half-light radii of Per~I, along with that of a field region of the same area, offset by $20'$ to the east of the dwarf galaxy. The CMD of the dwarf galaxy reveals a stellar overdensity below $i_\mathrm{P1}\simeq21.0$, the expected tip of the RGB (TRGB) of an old stellar population at the distance of M31. This overdensity is similar, but less pronounced than for Lac~I and Cas~III (Figure~2 of \citealt{martin13a}) and made more evident on the field-subtracted CMD displayed in Figure~\ref{LF}. The TRGB is well defined and will be used below to determine the heliocentric distance to the dwarf galaxy. Isolating these RGB-like stars reveals that they are clumped on the sky over a few arcminutes (right panel of Figure~\ref{CMD_map}), as expected for an M31 dwarf galaxy.

We now proceed to determine the structural parameters, distance, and total magnitude of Per~I based on the PS1 photometric catalogue. These properties are summarized in Table~1.

\begin{table}
\caption{\label{properties}Properties of Perseus~I/Andromeda~XXXIII}
\begin{tabular}{l|c}
\hline
$\alpha$ (ICRS) & $3^\mathrm{h}01^\mathrm{m}23.6^\mathrm{s}$\\
$\delta$ (ICRS) & $+40\deg59'18''$\\
$\ell$ ($\deg$) & 147.8\\
$b$ ($\deg$) & -15.5\\
$E(B-V)$\footnote{From \citet{schlafly11}.} & 0.115\\
$(m-M)_0$ & $24.49\pm0.18$\\
Heliocentric distance (kpc) & $785\pm65$\\
M31-centric distance (kpc) & $364^{+14}_{-10}$\\
$M_V$ & $-10.3\pm0.7$\\
$\mu_0$ (mag/arcsec$^2$) & $25.7^{+1.0}_{-0.9}$ \\
Ellipticity & $0.43^{+0.15}_{-0.17}$ \\
Position angle (N to E; $\deg$) & $-6^{+15}_{-10}$ \\
$r_h$ (arcmin) & $1.7\pm0.4$ \\
$r_h$ (pc) & $400^{+105}_{-85}$ \\
\end{tabular}
\end{table}

\subsection{Structure}
\label{structure}
\begin{figure}
\begin{center}
\includegraphics[width=0.65\hsize,angle=270]{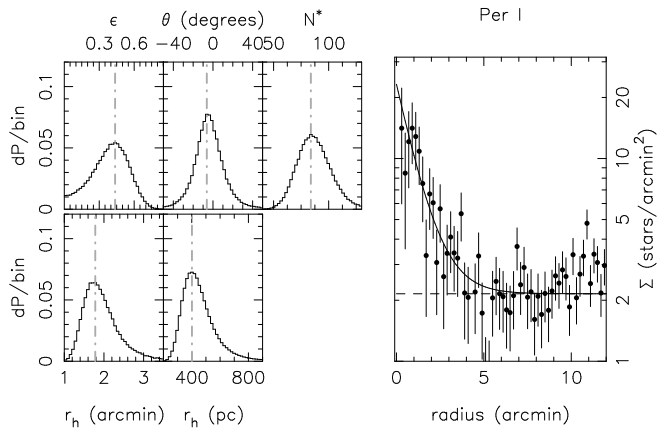}
\caption{\label{PerI_struct}\emph{Left panels: }The PDFs of Per~I's structural parameters, individually marginalized over the other four parameters. The maxima of these PDFs (gray dash-doted lines) are taken as the favored model parameters. \emph{Right panel: }The radial density profile of the dwarf galaxy, binned when assuming the favored centroid, ellipticity, and position angle (filled circles). The profile is compared to the favored exponential profile determined by the two other parameters ($r_h$ and $N^*$; full line), along with the implied background level (dashed line). The uncertainties on the binned data are assumed to be Poissonian.}
\end{center}
\end{figure}

Analogously to what we did for Lac~I and Cas~III, we determine the structural parameters of Per~I with an updated version of the maximum likelihood technique presented in \citet{martin08b}. The algorithm explores a 5-dimensional parameter space to calculate the likelihood of models of exponential radial density profiles with a constant background to represent the distribution of stars within $20'$ around Per~I. The parameters of the models are the centroid of the dwarf galaxy, its exponential half-light radius ($r_h$), the position angle of its major axis ($\theta$), its ellipticity ($\epsilon = 1-b/a$, with $a$ and $b$ being, respectively, the major and minor axis scale lengths of the system), and the number of member stars within the chosen CMD selection box ($N^*$).

The probability distribution functions (PDFs) returned for the most relevant parameters are displayed in Figure~\ref{PerI_struct}. In all cases, the parameters have well behaved PDFs and a favored model that maximizes the likelihood can easily be determined. The PDF for $N^*$ confirms that Per~I contains 2-3 times fewer detected stars than Lac~I or Cas~III (see \citealt{martin13a}), thereby explaining why it initially went undetected in our original search, and why its structural parameters have larger uncertainties. We nevertheless confirm that Per~I corresponds to a $>5$-sigma detection above the background. The quality of the fit is also represented in the right-hand panel of the figure by the very good agreement between the favored exponential profile, compared to the radial surface density profile of the stars, binned following the favored centroid, half-light radius, and ellipticity.

Per~I has a moderate ellipticity ($\epsilon=0.43^{+0.15}_{-0.17}$), and its size is $r_h=1.7\pm0.4$ arcmin, or $400^{+105}_{-85}\pc$ once we fold in the distance to the system. These values are quite typical for a dwarf galaxy.

\subsection{Distance}
\label{distance}
We use the TRGB, a good standard candle in the red (e.g. $i_\mathrm{P1}$), to determine the distance to Per~I. The probabilistic technique of \citet{makarov06} is applied under the assumption that Per~I is dominated by old stellar populations of intermediate to low metallicities. It yield a TRGB magnitude of $i_{\mathrm{P1},0}^\mathrm{TRGB} = 21.04\pm0.15$.

We determined in \citet{martin13a} that the absolute magnitude of the TRGB in the PS1 photometric system is $M_{i,\mathrm{P1}}^\mathrm{TRGB}=-3.45\pm0.10$, which yields the distance modulus of Per~I, $(m-M)_0 = 24.49\pm0.18$, corresponding to a heliocentric distance of $785\pm65\kpc$. This distance is statistically equal to M31's. It is therefore very likely that the new dwarf galaxy is a bound satellite of M31, even though this will have to be confirmed definitively from the radial velocity measurements we are currently gathering. From the angular separation of Per~I from M31, and the heliocentric distances of both systems, we derive an M31-centric distance of $364^{+14}_{-10}\kpc$, similar to And~XXVIII's distance on the other side of M31.

\subsection{Total magnitude}

We reproduce the analysis of \citet[][Section~3.3]{martin13a} to determine the total magnitude of Per~I: we start by summing up the flux contribution of all stars within the half-light radius of the dwarf galaxy that satisfy $20.7<i_{\mathrm{P1},0}<21.5$ and a color-cut to weed out obvious non-RGB stars. The expected flux contribution from foreground sources is calculated analogously, but over a larger field region. It is scaled to the area within Per~I's half-light radius and subtracted from the flux of the dwarf galaxy. The resulting contamination-corrected flux is in turn doubled to determine the total flux of the dwarf galaxy from detected PS1 stars. We then use the \textsc{Parsec} luminosity function of an old stellar population of intermediate metallicity ($12.7$ Gyr, $\FeH=-1.7$), selected in the PS1 photometric system \citep{bressan12}, to correct this flux for stars fainter than $i_{\mathrm{P1},0}=21.5$. This operation yields apparent magnitudes of $m_{r\mathrm{P1},0} = 15.0$ and $m_{i\mathrm{P1},0} = 14.5$.

With the PS1 $g_\mathrm{P1}$ observations barely grazing the TRGB of Per~I, we are forced to rely on a comparative analysis with other, well-known dwarf galaxies to transform these magnitudes to the commonly used total magnitude in the $V$ band. A comparison with And~I, II, III and~V, accounting for their distance moduli, eventually yields a total magnitude of $M_V = -10.3\pm0.7$ for Per~I.

\section{Concluding remarks}
In this letter, we present the discovery of a new Local Group galaxy, Perseus~I, in the stellar catalogue generated from stacked $r_\mathrm{P1}$- and $i_\mathrm{P1}$-PS1 imaging. The heliocentric distance to the new dwarf galaxy ($785\pm65\kpc$) is indistinguishable from that of M31. This makes Per~I a likely Andromeda companion, despite its sky position $27.9\deg$ east of Andromeda, which corresponds to a projected distance of $364^{+14}_{-10}\kpc$. The size of Per~I ($r_h=400^{+105}_{-85}\pc$) is typical for a dwarf galaxy of this luminosity ($M_V = -10.3\pm0.7$).

If confirmed as a companion of M31 by radial velocity measurements, the location of Per~I, far to the east of its host, makes it very useful to constrain the transverse motion of the M31 system as a whole, as exemplified by the analysis of \citet{vandermarel08}. If Per~I were moving within the Local Group with the motion of M31, the line-of-sight velocity vector of Per~I (which can be determined with an accuracy of better than $1\kms$) would hold about half of the transverse motion of M31. Of course a proper modelling of the satellite's orbital properties and anisotropy would need to be folded in for a reliable analysis. However, combining the information provided by And~VI,  And~XXVIII, Lac~I, and Per~I, all beyond $20\deg$ from M31, not obviously members of a co-moving group of satellites like LGS~3 and IC~10 \citep{ibata13a,tully13}, and spanning almost $60\deg$ on the sky, opens up very promising avenues for a measure of the M31 proper motion that is independent from its direct but arduous determination \citep{sohn12}.

Finally, the discovery of Per~I already brings the number of likely M31 satellite dwarf galaxies with $M_V<-6.5$ to at least 34 (30 dwarf spheroidal, 3 dwarf elliptical, 1 compact elliptical). Moreover, this number is expected to be a significant underestimation of the true number of M31 dwarf galaxies down to this magnitude since the fainter systems can currently only be found in PAndAS and dwarf galaxy detection limits are shallower in regions that are just covered by the SDSS or PS1. We nevertheless already know many more M31 than Milky Way dwarf galaxies, even if we include the fainter Milky Way systems discovered in the SDSS. This trend parallels that observed for globular clusters \citep[e.g.][]{caldwell11,huxor11}. Given that the number of dwarf galaxies orbiting a host correlates with their host's mass \citep[e.g.][]{starkenburg13}, it would be surprising if M31 were not significantly more massive than the Milky Way.

\acknowledgments

N.F.M. thanks Rodrigo Ibata for fruitful discussions and gratefully acknowledges the CNRS for support through PICS project PICS06183. N.F.M., E.F.S \& H.-W.R. acknowledge support by the DFG through the SFB 881 (A3). C.T.S. and E.F.B. acknowledge support from NSF grant AST 1008342.

The Pan-STARRS1 Surveys (PS1) have been made possible through contributions of the Institute for Astronomy, the University of Hawaii, the Pan-STARRS Project Office, the Max-Planck Society and its participating institutes, the Max Planck Institute for Astronomy, Heidelberg and the Max Planck Institute for Extraterrestrial Physics, Garching, The Johns Hopkins University, Durham University, the University of Edinburgh, Queen's University Belfast, the Harvard-Smithsonian Center for Astrophysics, the Las Cumbres Observatory Global Telescope Network Incorporated, the National Central University of Taiwan, the Space Telescope Science Institute, the National Aeronautics and Space Administration under Grant No. NNX08AR22G issued through the Planetary Science Division of the NASA Science Mission Directorate, the National Science Foundation under Grant No. AST-1238877, the University of Maryland, and Eotvos Lorand University (ELTE).



\end{document}